\documentstyle[prl,aps,epsfig,floats]{revtex}

\begin{document}
\title{Antiferromagnetic ordering in a 90 K copper oxide superconductor.}
\author{J.A. ~Hodges$^1$, Y. ~Sidis$^2$, P. ~Bourges$^2$, I. ~Mirebeau$^2$,
 M. Hennion$^2$ and X. ~Chaud$^3$.
\\
}
\address{
$^1$DRECAM-SPEC, CE-Saclay, 91191 Gif sur
Yvette, France.\\
$^2$ Laboratoire L\'eon Brillouin, CEA-CNRS, CE-Saclay, 91191 Gif sur
Yvette, France.\\
$^3$CRETA, CNRS, 25 Avenue des Martyrs, BP 166 38042 Grenoble cedex, France.\\
}
\date{\today}



\twocolumn[\hsize\textwidth\columnwidth\hsize\csname@twocolumnfalse\endcsname

\maketitle

\begin{abstract}
Using elastic neutron scattering, we evidence a commensurate antiferromagnetic
Cu(2) order (AF) in the superconducting (SC) high-$\rm T_c$ cuprate 
$\rm YBa_2(Cu_{1-y}Co_y)_3O_{7+\delta}$ (y=0.013, $\rm T_c$=93 K).  
However, the spin excitation spectrum is still dominated by a magnetic 
resonance peak at 41 meV  as in the Co-free system,  but with a 
reduced spectral weight. The substitution of Co thus leads to a
state where AF and SC cohabit. These 
results show that the hole-doped CuO$_2$ plane is close to an AF instability 
even when T$_c$ remains optimum.

\end{abstract}


]

\narrowtext

The interplay between magnetic order and superconductivity is an 
interesting and profound phenomenon ubiquitous in strongly correlated systems, 
such as high-$\rm T_c$ cuprates, low-$\rm T_c$ ruthenates and heavy fermions  
systems. 
There have been a number of reports of the
coexistence of magnetic order and exotic superconductivity: 
for example, in Ce and U-based heavy fermion systems\cite{Mathure98,Amato97}, 
in superoxygenated $\rm La_2CuO_{4+\delta}$,  
in $\rm La_{1.6-x}Nd_{0.4}Sr_xCuO_4$ 
\cite{Lee99,Tranquada99} and more recently  
in the well underdoped regime of $\rm YBa_2Cu_3O_{6+x}$  (x=0.5-0.6)  
\cite{Sidis01,Mook01}. 
In the last case \cite{Mook01}, the observation that the magnetic intensity decreases at large
momentum more rapidly than does the Cu-spin form factor, has been  considered to support 
the suggestion  of orbital moments of a d-wave density-wave (DDW) order parameter 
(that could be the hidden order responsible 
for the "pseudo-gap phase" of underdoped cuprates) \cite{CLMN}.
However, in the cuprates, the real coexistence of  
magnetic order and superconductivity or a microscopic phase segregation 
remains a matter of discussion. Especially, the role played by disorder is 
still an open  
question \cite{Kohno99}. In this letter, we evidence the appearance of
antiferromagnetic order in the ${\rm CuO_2}$ 
planes of a fully oxydised superconducting $\rm YBa_2Cu_3O_{7}$ based
system with ${\rm T_c}$=93 K, when a disorder is introduced through the  
substitution of cobalt atoms at the copper site of the chains.

The $\rm YBa_2Cu_3O_{6+x}$ perovskite structure contains two copper sites:  
Cu(1) belonging to the Cu-O chains (along the {\bf b}-axis), and Cu(2) 
belonging  
to the $\rm CuO_2$ planes. $\rm Co^{3+}$ ions substitute only at  
the Cu(1) sites \cite{Tarascon88}. Due to its higher oxidation state  
compared to that of Cu(1), the Co cation pulls in extra oxygen to 
increase its  oxygen-coordination. Each added Co atom, with an  
average coordination number of 5, pulls in  0.5 oxygen atoms  
\cite{Tarascon88,Howland89,xafs,Renevier94}. Additionally, the Co substitution
induces a transverse distortion of its Cu(1) site \cite{xafs,Renevier94}. 
Co atoms tend to form either small  
clusters like dimers \cite{Tarascon88,Renevier94} or even short chains  
along the (110) direction \cite{xafs}. As a result, these chains pin down  
the twin boundaries of the orthorhombic structure (micro-twinning) 
that triggers an orthorhombic-tetragonal transition for y $\ge$ 0.025  
\cite{schmahl}. NMR\cite{dupree} and transport measurements
\cite{Clayhold89} show the cobalt substitution reduces the hole doping.
However, for the low Co substitution level (y $\simeq$ 0.013) examined 
here, despite the decrease in the Hall effect derived 
carrier density \cite{Clayhold89} and in the specific heat derived condensate 
density \cite{Loram91},
the doping is still high enough for ${\rm T_c}$ to remain at its optimum 
value. In addition, Nuclear Quadrupole Resonance (NQR)
\cite{Matsumura94} 
measurements have evidenced the appearance of magnetic moments on the Cu of 
the Cu(1) sites and more surprisingly on those of the Cu(2) sites as well. 
M\"ossbauer probe measurements 
\cite{Hodges} have also evidenced the moments on the Cu(2) sites.

\begin{figure}[t]
\epsfxsize=9cm
$$
\epsfbox{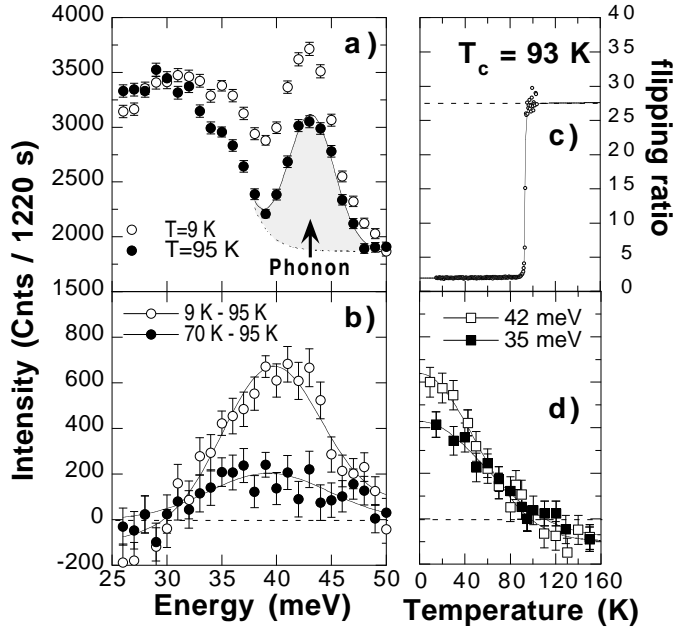}  
$$
\caption[xxx]{ 
a) Energy scans performed at {\bf Q}=(0.5,0.5,-5.4) at T=9.2 K and  
T=95 K. The scan above ${\rm T_c}$ looks very similar to that 
in the Co-free sample with the phonon peak at 42.5 meV\cite{bourges96}.  
b) Difference between energy scans performed at 9.2 K (or 70 K) and 
at 95 K. c) determination of $\rm T_c = 93$ K by the neutron depolarisation 
technique. d) temperature dependence of the AF response at 35 meV and 
42 meV.}
\label{Escan}
\end{figure}

We present a neutron scattering study of the magnetic properties of a large 
single crystal (1.4 $\rm cm^{3}$) of fully oxygenated 
$\rm YBa_2(Cu_{1-y}Co_y)_3O_{7+\delta}$. The sample was prepared by the 
top-seed melt texturing method. A microprobe analysis confirmed the Co 
content was that of the starting mixture, y=0.013, and 
showed the Co was uniformly distributed on the $\mu$m scale.
Neutron depolarization measurements (Fig \ref{Escan}.c) 
provided $\rm T_c$= 93 K. 

The neutron scattering experiments were performed on the triple axis 
spectrometers 1T1 and 4F2 at the Laboratoire L\'eon Brillouin, Saclay (France).
For elastic neutron scattering measurements (ENS) on 4F2,
double PG(002) monochromators and analyzer were used and a
beryllium filter was inserted into the scattered beam in order to
remove higher order contamination. The data were taken with a fixed final
wavevector of 1.55 \AA$\rm ^{-1}$.
For the inelastic neutron scattering measurements (INS) on 1T1,
a focusing Cu(110) monochromator and  a PG(002) analyzer were used and a
pyrolytic graphite filter was inserted into the scattered beam.
The data were taken with a fixed final
wavevector of 4.1 \AA$\rm ^{-1}$. Measurements were carried out with the crystal 
in two different orientations where wave vector transfers of the form 
{\bf Q}=$\rm (H,H,L)$ and $\rm (3H,H,L)$, respectively, were accessible.
Throughout this article, the wave vector {\bf 
Q} is indexed in units of the reciprocal tetragonal lattice vectors 
2$\pi$/a=2$\pi$/b=1.63 \AA$\rm ^{-1}$
 and 2$\pi$/c=0.53 \AA$^{-1}$. In this notation
the $\rm (\pi/a,\pi/a)$ wave vector parallel to the ${\rm Cu O_2}$ planes
corresponds to points of the form (h/2,k/2) with h and k odd integers.

\begin{figure}[t]
\epsfxsize=7.7cm
$$
\epsfbox{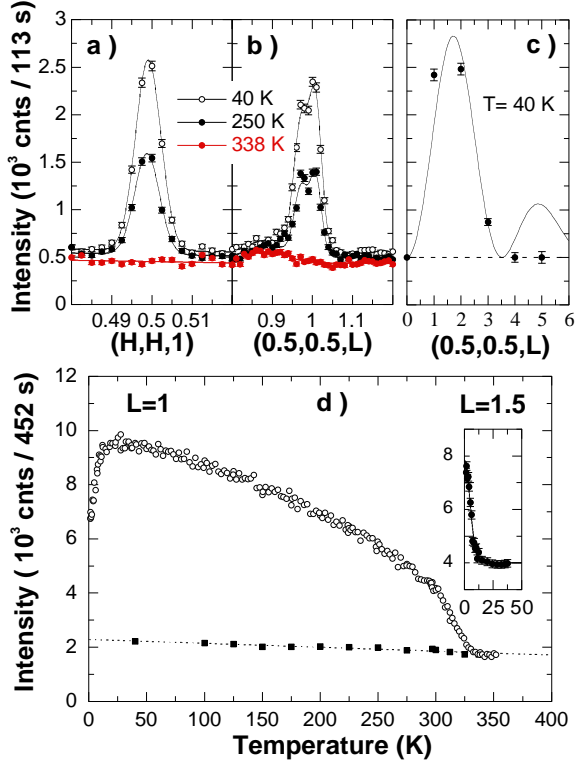}  
$$
\caption{ 
Elastic neutron intensity along a) the (110) direction 
and b) the (001) direction around {\bf Q}=(0.5,0.5,1). 
c) L-dependence of magnetic intensity at different AF peaks, 
{\bf Q}=(0.5,0.5,L). The full line represents the magnetic intensity 
expected from in-plane Cu(2) spins with isotropic Cu$^{2+}$ form factor 
and including the resolution correction. 
d) Temperature dependences of the neutron scattering intensity at
{\bf Q}=(0.5,0.5,1). The full squares represent the background from 
scans as shown in a) and b). Inset of d) shows the appearance of magnetic  
intensity at {\bf Q}=(0.5,0.5,1.5) below $T_m \simeq$ 12 K where the  
peak at {\bf Q}=(0.5,0.5,1) displays a re-entrant behavior. 
}
\label{elastic}
\end{figure}

We first describe the observation of the commensurate magnetic order. 
Figure \ref{elastic} shows  
the elastic neutron intensity at the antiferromagnetic wavevector  
{\bf Q}=(0.5,0.5,1) along both the (110) and (001) directions.  The peak, 
absent at 338 K and present at 250 K and 40 K, evidences the magnetic order.
Different antiferromagnetic peaks, {\bf Q}=(0.5,0.5,L) with L integer, are 
reported in  Fig. \ref{elastic}.c where magnetic intensity is sizeable at 
L=1,2,3. 
The observed pattern (Fig. \ref{elastic}.c) and 
the absence of any magnetic peak at  {\bf Q}=(0.5,0.5,0) implies that the 
magnetic response is fully dominated by magnetic moments at the Cu(2) sites.
Assuming all of the Cu(2) carry 
the same ordered moment, we obtain a low temperature (T=40 K) mean moment of 
$\sim 0.10 \pm 0.05 \mu_B$.
Gaussian fits of Fig. \ref{elastic} at different temperatures shows that the
AF order is resolution limited, meaning that the correlation
lengths are typically $\xi >$ 200 \AA. 
A detailed q-dependence of the magnetic peak, see  for instance the 
double peak structure of the scan along {\bf c$^*$} of Fig. \ref{elastic}.b,
shows a mosaic distribution of the magnetic peak which does 
not exactly reproduce that of the nuclear peaks. This means that the 
volume of AF region does not exactly match the volume of the single crystal.

The temperature dependence of the neutron scattering intensity measured at  
the antiferromagnetic wavevector {\bf Q}=(0.5,0.5,1) (Fig. \ref{elastic}.d) 
shows the system orders at $T_N \sim$ 330 K. As the temperature is lowered,
the AF Bragg intensity initially increases continuously and no anomaly is 
observed on passing through $\rm T_c$. The peak intensity
displays a marked downturn at $\rm T_m \simeq$ 12 K, and almost half of 
its intensity is left as T$\rightarrow$ 0. Below $\rm T_m$, additional neutron 
intensity occurs at L=1.5 indicating the system undergoes an 
AFI-AFII transition, characterized by the doubling of the AF unit cell along 
the c axis.  This transition is also observed in $\rm YBa_2Cu_3O_6$ when 
substituted at the Cu(1) site\cite{mirebeau,casalta} and AFII ordering 
is observed in non-superconducting $\rm YBa_2(Cu_{1-y}Co_y)_3O_{7+\delta}$ 
with high Co substitution levels \cite{Zolliker88}. 
The re-ordering observed in the present case may
be linked to the influence of magnetic freezing which is known to exist in
in similar samples,
For example, in $\rm YBa_2(Cu_{0.94}Co_{0.06})_3O_{7+\delta}$, 
time of flight neutron scattering measurements \cite{bellouard} have 
evidenced the progressive freezing of the Co moments as the temperature is 
lowered in agreement with  local probe measurements \cite{Matsumura94,Hodges}.

Without producing any reduction in 
the superconducting transition temperature, the substitution of magnetic 
Co$^{3+}$ 
at the chain sites thus introduces a very specific perturbation which induces 
a commensurate AF 3D order, below T$_N$ $\sim$ 330 K, at the copper sites of 
the CuO$_2$ planes. 
A key question is how does the superconductivity and the antiferromagnetic 
order cohabit ?

To address this issue, we first consider the possibility of complete phase
segregation, such that 
only a small fraction of Cu(2) atoms, for example those adjacent to a Co atom,
carries the full magnetic moment of the undoped 
cuprates (0.6 $\mu_B$). The concentration of Cu(2) atoms (in the planes) 
needed to account for the observed scattering intensity would then 
be $\sim$ 3 \% which is roughly comparable to the Co substitution level 
(in the chains). There are a number of arguments against this hypothesis.
As the magnetic correlation length exceeds 200 \AA, then
the Co would be  essentially concentrated within these clusters. Such Co clusters would
yield an observable magnetic diffraction pattern. The local Co 
concentration would also be quite high and we would expect to observe the behavior 
seen in samples having high Co levels. For example,
insulating samples of $\rm YBa_2(Cu_{1-y}(Co,Fe)_y)_3O_{7+\delta}$ 
show the AFII structure up to T$_N \sim$ 400 K
\cite{mirebeau,Zolliker88}, whereas it is the AFI structure that is observed here at T$_N$.
Neither of these features is observed.
Further arguments against complete phase separation is provided by
Cu-NQR \cite{Matsumura94} or Y-NMR \cite{dupree} local probe 
measurements in $\rm YBa_2(Cu_{1-y}Co_y)_3O_{7+\delta}$ which evidence features
that are different from those of the undoped insulating AF state. We recall 
also that local probe Cu NQR \cite{Matsumura94} and M\"ossbauer \cite{Hodges} 
measurements on samples having the same low Co substitution level as the 
sample studied here have shown that over 50 \% of the Cu(2) carry magnetic 
moments. Thus, it does not seem possible that the AF order 
could be linked to the existence of locally undoped regions. Consequently, 
the two phenomena (AF and SC) appear to be in contact at a microscopic level.

As the Co atoms aggregate into dimers or small clusters  
forming lines along the (110) direction
\cite{xafs}, we speculate that these lines of magnetic Co atoms are the 
perturbing elements which induce the AF order into the CuO$_2$ planes. 
However, we recall the Cu(2) moments are not confined to the immediate 
vicinity of these lines for the magnetic correlation lengths greatly exceed 
the lateral dimensions of a twin boundary. 

Finally, we believe this new AF order is analogous to that recently 
reported in underdoped $\rm YBa_2Cu_3O_{6+x}$ ($x \sim$ 0.5 - 0.6) 
\cite{Sidis01,Mook01} with however one striking difference: the additional 
enhancement of the AF intensity observed below ${\rm T_c}$ is not seen 
in the present case. At large Q, the intensity decreases much more 
rapidly than expected for the Cu-spin form factor (see Fig. \ref{elastic}.c). 
This result is similar to that observed in well underdoped 
YBa$_2$Cu$_3$O$_{6.6}$\cite{Mook01} where it was interpreted as evidencing 
DDW order\cite{CLMN}. Our finding of a similar unusual structure factor in 
a Co-substituted sample which is near optimal doping and where a pseudo-gap 
behavior is absent, questions this conclusion. The precise reason for the 
observed structure factor remains unclear at present.

\begin{figure}[t]
\epsfxsize=5.5cm
$$
\epsfbox{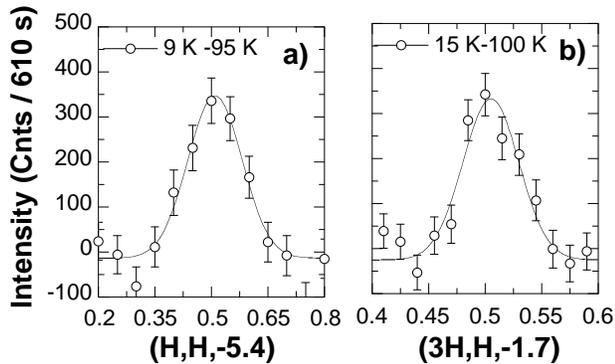} 
$$
\caption{ 
Difference of constant energy scans performed at 39 meV measured at low temperature,
and just above $\rm T_c$: a) around {\bf Q}=(0.5,0.5,-5.4) 
along the (110) direction: b)around {\bf Q}=(1.5,0.5,-1.7) along the (310) direction. 
 Solid lines are fit by Gaussian lineshape.}
\label{Qscan}
\end{figure}

We next present the inelastic magnetic fluctuations around the AF wavevector.
In cobalt-free optimally doped $\rm YBa_2Cu_3O_{7-\delta}$, the AF 
correlations are  
purely dynamic, and the spin excitation spectrum in the superconducting 
state is characterized by a sharp antiferromagnetic excitation peaked at  
41 meV, the so-called "magnetic resonance peak" \cite{rossat91,bourges96}. 
In our 
cobalt substituted sample, we looked for this magnetic excitation specific of 
$d$-wave superconductivity and we performed constant energy scans at 
39 meV around
Q=(0.5,0.5,-5.4) along the (110) direction as well as around (1.5,0.5,-1.7)
along the (310) direction (Fig.~\ref{Qscan}). At low temperature, a peak
shows up in both scans centered at the AF wave vector. The ratio of the 
intensity of both scans evolves as a function of Q as expected for the 
Cu$^{2+}$ anisotropic magnetic form factor \cite{casalta}.  In both scans, the
peak diminishes drastically at $\rm T_c$. Above $\rm T_c$, weaker intensity, 
peaked at AF wavevector, remains in both scans, in agreement with the 
Co-free compound $\rm YBa_2Cu_3O_{6.97}$\cite{bourges96}. 
After subtraction of the scan just above $\rm T_c$ from that at low 
temperature, the remaining intensity was fitted to a Gaussian profile centered
at the AF wave vector. For both types of constant energy scans reported in 
Fig.~\ref{Qscan}, the AF response at 39 meV in the superconducting state 
displays a momentum width 
(FWHM) of 0.28 $\pm$ 0.06 \AA$\rm ^{-1}$. This momentum distribution
is similar to that of the magnetic resonance peak in the Co-free system 
\cite{bourges96}.


Figure ~\ref{Escan} shows energy scans performed at Q=(0.5,0.5,-5.4) at
low temperature and just above $\rm T_c$. The enhancement of the AF response
around 41 meV (the magnetic resonance peak) is visible
in the raw data and it is further confirmed by the differences shown on
Fig.~\ref{Escan}.b. The magnetic resonance peak is not sharp in energy and can be fitted 
by  the usual "ansatz" of a single Gaussian lineshape: this yields an 
intrinsic energy width of $\sim$ 9 $\pm$ 1 meV (FWHM). 
This analysis in terms of a single broad signal centered at $\rm E_r$
is supported by the temperature dependences performed at the AF wave vector at 
42 meV and 35 meV (6 meV below $\rm E_r$) which both show a similar decrease of 
the AF response up to $\rm T_c$ (Fig.~\ref{Escan}.c). As a function of  
temperature, the magnetic resonance peak disappears at $\rm T_c$
without any significant shift of its characteristic energy (Fig.~\ref{Escan}.b).
Furthermore, its energy and momentum integrated intensity, calibrated in absolute unit 
against the phonon at 42.5 meV, is $\sim$ 0.025 $\rm \mu_B^2$, i.e it is 
about twice  
 weaker than  that reported in the Co-free compound \cite{fong00}.
 
The spin dynamics observed following Co substitution (at the Cu(1) site) show
some features which are common to those observed with Ni and Zn  
substitutions (at the Cu(2) site), where a broadened  
magnetic resonance peak was also seen \cite{sidis00}. 
However, the Ni and Zn substitutions have more drastic effects on the 
dynamical AF correlations: Zn already induces strong AF fluctuations in the 
normal state although it reduces the enhancement of the spin susceptibility in
the SC state, whereas Ni renormalizes the resonance energy. 


Following the observed
50$\%$ reduction in the weight of the resonance peak, it is 
possible that distinct superconducting regions occupy about half of the
sample volume leaving the remainder for the distinct magnetically ordered 
regions. This fully agrees with NQR results on  sample with similar Co content
which show that about half the Cu(2) carry magnetic moments 
\cite{Matsumura94}.
Such a phase segregation scenario has some analogies with the {\it three phase 
model} developed to explain the Cu-NQR data\cite{Matsumura94}. 
However, the magnetic correlation lengths found here 
($\xi > 200\AA$) are much bigger than the dimensions of the nucleated regions 
where the Cu(2) become magnetic as considered in ref.\cite{Matsumura94}.
In addition, Kohno  {\it et al.} \cite{Kohno99} have proposed a 
phenomenological description of AF-SC coexistent states due to disorder in 
strongly correlated systems. In their model, based on Ginsburg-Landau theory, 
a key assumption is that the SC state is in competition with the AF phase 
with a first order phase boundary, which enables the AF state to nucleate 
where SC is suppressed. For a finite concentration of impurities, the first 
order AF-SC boundary of the clean case is replaced by a finite region where 
the SC and AF moments coexist microscopically with spatially varying order 
parameters. 
It has also been proposed 
\cite{ohashi} that an AF state can locally appear 
around surfaces or impurities and can co-exist with $d$-wave superconductivity
and in particular, the local formation of an AF order parameter can easily 
occur near a (110) surface.
The AF order in the present case could have its origin in the magnetic 
polarisation produced by the Co$^{3+}$ or in the structural effects related to 
the micro-twinning along the (110) directions \cite{schmahl} or more probably,
to a combination of both these effects. 

As a conclusion, we observe a Cu(2) site commensurate long range
AF order in the superconducting high-$\rm T_c$ cuprate 
$\rm YBa_2(Cu_{1-y}Co_y)_3O_{7+ \delta}$ (y=0.013, $\rm T_c$=93 K).
The observed structure factors differ slightly from that for the
Cu(2) spins in the undoped state of the cuprates. This difference, observed in
a sample where the doping level is high enough to support an optimum T$_c$ 
value, questions the conclusion that a similar observation in well underdoped
YBa$_2$Cu$_3$O$_{6.6}$ is evidence for the predicted DDW order in the
pseudo-gap phase.
The cohabitation of AF and SC  can be described as a formation of an AF phase 
within a $d$-wave superconductor.  Whatever the precise mechanism giving rise
to the AF ordering, our results reveal that hole-doped CuO$_2$ plane is 
close to an AF instability even when T$_c$ remains optimum.

We wish to thank Pierre Gautier-Picard and Philippe Mendels for fruitful 
discussions.

\end{document}